\documentclass{elsart41}
\usepackage{graphics}
\usepackage{graphicx}
\usepackage{amssymb}
\usepackage{amsmath}
\begin{document}

\begin{frontmatter}

% Title, authors and addresses

% use the thanksref command within \title, \author or \address for footnotes;
% use the corauthref command within \author for corresponding author footnotes;
% use the ead command for the email address,
% and the form \ead[url] for the home page:
% \title{Title\thanksref{label1}}
% \thanks[label1]{}
% \author{Name\corauthref{cor1}\thanksref{label2}}
% \ead{email address}
% \ead[url]{home page}
% \thanks[label2]{}
% \corauth[cor1]{}
% \address{Address\thanksref{label3}}
% \thanks[label3]{}

\title{Proposal for a two-channel quantum dot set-up: 
 Prediction for the capacitance lineshape}
%---- Don't remove this comment line! ----
%
% use optional labels to link authors explicitly to addresses:
% \author[label1,label2]{}
% \address[label1]{}
% \address[label2]{}

\author[DE]{N. Shah\corauthref{Shah}},
\ead{nayana@thp.uni-koeln.de}
\author[CH]{C. J. Bolech}

\address[DE]{Institut f\"ur Theoretische Physik, Universit\"at zu
  K\"oln, Z\"ulpicher Strasse 77, D-50937, K\"oln, Germany}
\address[CH]{Universit\'{e} de Gen\`{e}ve, DPMC, Quai Ernest Ansermet 24, CH-1211 Gen\`{e}ve 4, Suisse}

\corauth[Shah]{Corresponding author. Tel: +49 221 470 4309 Fax:
+49 221 470 2189}

\begin{abstract}

We have made a detailed proposal for a two-channel quantum dot set-up. The energy scales in the problem are such that we are able to make connection with the two-channel Anderson
model, which, in spite of being well-known in the context of heavy-Fermion systems remained theoretically elusive until recently and lacked a mesoscopic realization. Verification of our precise and robust predictions for the differential
capacitance lineshape of the dot will provide an experimental signature of the two-channel behavior.

\end{abstract}

\end{frontmatter}

% main text

The physics of quantum impurities continues to be one of the
central themes in the study of electronic systems exhibiting
strong correlations. In the subject's long history, one of the
recent developments --taking place during the last decade-- was the
realization of the Kondo effect in artificially fabricated
nanostructures \cite{ref1}. The appeal of this new type of quantum
impurities resides in the high degree of experimental control over
the parameters of the system and the possibilities to carry out
new --more direct-- types of measurements. Because of this unique
characteristics, artificial structures like quantum dots are very
promising for the experimental realization of the elusive
two-channel Kondo effect \cite{ref2}.

\begin{figure}[!ht]
\begin{center}
\includegraphics[width=0.45\textwidth]{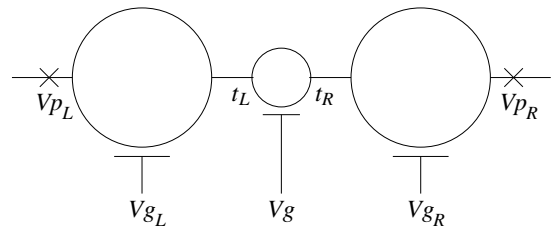}
\end{center}
\caption{Schematic depiction of the set-up involving two isolated
large quantum dots (grains) independently connected to a small
quantum dot located between them. The tunneling barriers between
the grains and the central dot, $t_L$ and $t_R$, can be
independently adjusted tuning the voltages $V_{t_L}$ and
$V_{t_R}$. The occupancy of the dot and the grains are controlled
by the respective gate voltages: $Vg$, $Vg_L$, and $Vg_R$. The
dot-grains system is isolated from the two-dimensional electron
gas surrounding it, so that the total charge is conserved. The
plunger gates, $Vp_L$ and $Vp_R$, are completely closed once the
system is tuned to the triple charge-degeneracy point of dot and
grains.} \label{fig1}
\end{figure}

In a recent work \cite{ref3}, we have proposed a new set-up that
realizes two-channel Kondo physics. Even more, the set-up is
described precisely by the two-channel Anderson model, which is a
well known model in the context of heavy fermion physics
\cite{ref4}. A lot of progress was made during recent years in
unraveling the physics of two-channel Anderson impurities
\cite{ref5,ref6}, what allows us to make precise predictions for
the quantum-dot set-up that we propose. The set-up, illustrated in
Figure \ref{fig1}, consists of a small quantum dot sided by two
large Coulomb-blockaded mesoscopic islands that we shall call
\emph{grains}.

A study of Coulomb energy of the different electronic
configurations shows that the medium and low temperature physics
of the dot-grains system is captured by the two-channel Anderson
model \cite{ref3}
\begin{align}
\hat{H} &
=\sum\nolimits_{\alpha}\hat{H}_{\mathrm{grain}}^{\alpha}+\hat
{H}_{\mathrm{conf}}+\hat{H}_{\mathrm{tun}}\label{eqn: hamil}\\
\hat{H}_{\mathrm{grain}}^{\alpha} &
=\sum\nolimits_{k\sigma}\varepsilon
_{k\sigma}^{\alpha}\,\hat{g}_{k\alpha\sigma}^{\dagger}\hat{g}_{k\alpha\sigma
}^{%
%TCIMACRO{\TeXButton{t}{\phantom{\dagger}}}%
%BeginExpansion
\phantom{\dagger}%
%EndExpansion
}\\ \hat{H}_{\mathrm{conf}} &
=\sum\nolimits_{\sigma}\varepsilon_{f}\,\hat
{f}_{\sigma}^{\dagger}\hat{f}_{\sigma}^{%
%TCIMACRO{\TeXButton{t}{\phantom{\dagger}}}%
%BeginExpansion
\phantom{\dagger}%
%EndExpansion
}+\sum\nolimits_{\alpha}\varepsilon_{b\alpha}\,\hat{b}_{\bar{\alpha}}%
^{\dagger}\hat{b}_{\bar{\alpha}}^{%
%TCIMACRO{\TeXButton{t}{\phantom{\dagger}}}%
%BeginExpansion
\phantom{\dagger}%
%EndExpansion
}\\ \hat{H}_{\mathrm{tun}} &
=\sum\nolimits_{k\alpha\sigma}t_{k\alpha}\left[
\hat{g}_{k\alpha\sigma}^{\dagger}\hat{b}_{\bar{\alpha}}^{\dagger}\hat
{f}_{\sigma}^{%
%TCIMACRO{\TeXButton{t}{\phantom{\dagger}}}%
%BeginExpansion
\phantom{\dagger}%
%EndExpansion
}+\hat{f}_{\sigma}^{\dagger}\hat{b}_{\bar{\alpha}}^{%
%TCIMACRO{\TeXButton{t}{\phantom{\dagger}}}%
%BeginExpansion
\phantom{\dagger}%
%EndExpansion
}\hat{g}_{k\alpha\sigma}^{%
%TCIMACRO{\TeXButton{t}{\phantom{\dagger}}}%
%BeginExpansion
\phantom{\dagger}%
%EndExpansion
}\right]
\end{align}
The notation is as follows: $g_{k\alpha\sigma}$ are the electron
operators for the two grains (labeled with $\alpha$) and
$b_{\bar{\alpha}}$ and $f_\sigma$ are, respectively, boson and 
fermion fields describing the different low-energy
Coulomb-blockade configurations. The charge fluctuations between
dot and grains are described by the term $\hat{H}_{\mathrm{tun}}$. The last two terms
in the Hamiltonian together with the constraint
\begin{equation}
\sum\nolimits_{\sigma}\hat{f}_{\sigma}^{\dagger}\hat{f}_{\sigma}^{%
%TCIMACRO{\TeXButton{t}{\phantom{\dagger}}}%
%BeginExpansion
\phantom{\dagger}%
%EndExpansion
}+\sum\nolimits_{\alpha}\hat{b}_{\bar{\alpha}}^{\dagger}\hat{b}_{\bar{\alpha}%
}^{%
%TCIMACRO{\TeXButton{t}{\phantom{\dagger}}}%
%BeginExpansion
\phantom{\dagger}%
%EndExpansion
}=1
\end{equation}
encode the physics of the Coulomb blockade. The different
parameters in the Hamiltonian are directly related to the
corresponding gate voltages shown in Figure \ref{fig1}, and can be
tuned externally in order to explore the different regimes of the
system.

The nature of the set-up we propose precludes the possibility of
transport measurements as it stands. On the other hand, a feasible and very
interesting experiment would be to measure the
capacitance lineshape of the central dot. We have recently
explained in detail how that can be achieved using a system of two
symmetric single electron transistors attached one to each grain.
The differential capacitance probes directly the charge
susceptibility of the system, which one can compute from the
two-channel Anderson Hamiltonian using a Thermodynamic Bethe
Ansatz analysis \cite{ref5}. Namely, $\delta
C\propto\chi_c=-\partial Q/\partial\epsilon$ and in the low
temperature limit, the dot excess charge $Q\!\equiv e\!\sum\nolimits_{\sigma
}\!\!\!<\!\!\hat{f}_{\sigma}^{\dagger}\hat{f}_{\sigma}^{%
%TCIMACRO{\TeXButton{t}{\phantom{\dagger}}}%
%BeginExpansion
\phantom{\dagger}%
%EndExpansion
}\!\!>$ is
\[
Q =\int_{-\infty}^{+\infty}\int_{-\infty}^{0}%
\frac{\left(  2z/\pi\right)  ~dz}{\cosh\left[
\frac{\pi}{2\Delta}\left( x-z\right)  \right]  }\frac{\left(
x-\varepsilon\right)  ~dx}{\left[  \left(
x-\varepsilon\right)  ^{2}+4\Delta^{2}\right]  ^{2}}%
\]

\begin{figure}
[t]
\begin{center}
\includegraphics[
height=2.3333in,
width=3.3278in
]%
{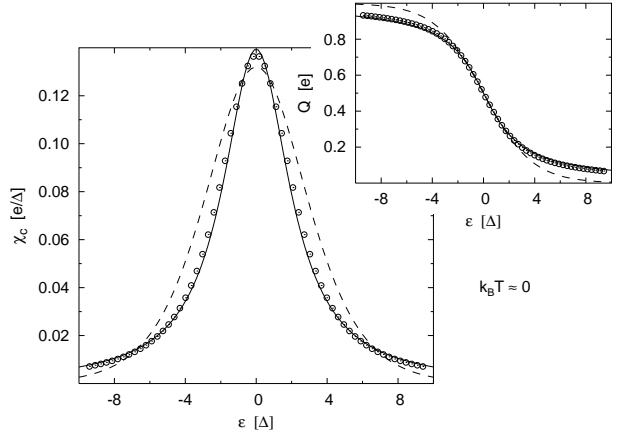}%
\caption{Charge susceptibility of the quantum dot as a function of energy
splitting. The open circles indicate the vanishing-temperature TBA result. The
solid line corresponds to a Lorentzian fit, while the dashed line is the fit
to the derivative of a Fermi-Dirac distribution. With the same symbols as
before, the overlay shows the excess charge on the dot as a function of energy
splitting.}%
\label{fig2}%
\end{center}
\end{figure}

Figure \ref{fig2} plots the vanishing-temperature $\chi_c$ and the overlay shows $Q$, both as a function of $\epsilon$. We have argued \cite{ref3} that this lineshape is different from
the one expected in other proposals for two-channel Kondo
behavior and different as well from what one would expect for a
one-channel system. The functional form of the lineshape is also
robust to changes in temperature provided $T\lesssim\Delta\propto
t_{L,R}^2$, as well as to the presence of magnetic field or grain
asymmetries bounded also by $\Delta$. Notice that this prediction
has no free fitting parameters once the value of $\Delta$ is
determined. We expect that the detailed information available for
the system from the theoretical point of view might act as an
encouragement for the experimental groups able to test our
predictions.

\section*{Acknowledgement}
Part of this work was supported by the Swiss National Science
Foundation within the MaNEP federal initiative.

\end{document}